\documentclass[11pt]{article}
\pdfoutput=1
\usepackage{aas_macros,amsmath,amssymb,comment,cite,esint,graphicx,mathtools}
\usepackage[margin=.8in,letterpaper]{geometry}
\usepackage[colorlinks=true]{hyperref}
\usepackage[affil-it]{authblk}
\usepackage{subcaption}
\usepackage[utf8]{inputenc}
\usepackage{mathrsfs}
\usepackage{appendix}
\usepackage{amssymb}
\usepackage{float}
\usepackage{physics}
\usepackage{color}
\usepackage{cite}
\usepackage{hyperref}
\hypersetup{pageanchor=false}
\usepackage{indentfirst}
\usepackage{url}
\usepackage{caption}
\usepackage[numbers,square,comma,sort&compress,merge]{natbib}
\usepackage{esint}
\usepackage{overpic}
\usepackage{graphicx}
\usepackage{epsf,amsmath,bbold,amsfonts,stmaryrd}
\usepackage{textcomp}
\usepackage{ulem}
\usepackage{tikz}
\usepackage{amsfonts}
\usepackage{array}
\usepackage{longtable}
\usepackage[utf8]{inputenc}
\usepackage{multirow}
\usepackage{orcidlink}
\numberwithin{equation}{section}
\let\originalleft\left
\let\originalright\right
\renewcommand{\left}{\mathopen{}\mathclose\bgroup\originalleft}
\renewcommand{\right}{\aftergroup\egroup\originalright}
\mathcode`\*="8000
{\catcode`\*=\active\gdef*{\mathclose{}\,\mathopen{}}}

\newcommand{\be}{\begin{equation}}
\newcommand{\ee}{\end{equation}}
\newcommand{\bea}{\setlength\arraycolsep{2pt} \begin{eqnarray}}
\newcommand{\eea}{\end{eqnarray}}

\def\be{\begin{equation}}
\def\ee{\end{equation}}
\def\bag{\begin{aligned}}
\def\eag{\end{aligned}}
\def\bea{\begin{eqnarray}}
\def\eea{\end{eqnarray}}
\def\ba{\begin{array}}
\def\ea{\end{array}}

\def\bc{\begin{center}}
\def\ec{\end{center}}

\usepackage{booktabs}
\usepackage{multirow}
\usepackage{amsmath} 

\begin{document}
\title{The extended inner shadow of Kerr-Taub-NUT black hole with thin disk flows}

\author{Yu-Yan Wang$^1$$\textsuperscript{\dag}$~\orcidlink{0009-0008-3382-8425}, 
Meng-Die Zhao$^1$$\textsuperscript{\ddag}$~\orcidlink{0009-0009-6644-1777}, 
Xin-Yu Wang$^2$\textsuperscript{\S}~\orcidlink{0000-0002-9599-4743}
 and 
Guo-Ping Li $^1$$\textsuperscript{*}$~\orcidlink{0000-0002-2788-9861}}

\date{}

  \maketitle

\vspace{-10mm}
\begin{center}
\textit{
$^1$ School of Physics and Astronomy, China West Normal University, Nanchong 637000, China\\
$^2$ School of physics and astronomy, Beijing Normal University, Beijing 100875,China }
\end{center}

\vspace{8mm}

\begin{abstract}
In this paper, we apply numerical backward ray-tracing to study the observational appearance of Kerr--Taub--NUT (KTN) black holes illuminated by thin accretion disk flows. We obtained the inner shadow, redshift characteristics, and intensity distribution of thin-disk images of the KTN black hole, as observed by a common observer located at different positions. The results show that increasing the spin parameter progressively deforms the critical curve into a ``D'' shape while simultaneously shrinking and distorting the inner shadow. More importantly, for $n=0.3$ at $\theta_o=80^\circ$, the inner shadow develops a novel ``duck-cap-like'' morphology with a sharply protruding lower-right edge beyond the critical curve. We term this feature the ``extended inner shadow'', a structure distinct from the Kerr case. Unlike the standard inner shadow, it consists partly of photons absorbed by the horizon and partly of photons that avoid both absorption and crossing the disk plane, thus receiving no emission. Such deviations from Kerr predictions could be tested by future high-precision astronomical observations, potentially offering new evidence for the existence of NUT charge(or the gravitomagnetic monopole) in black hole.

\end{abstract}

\vfill
\begin{flushleft}    
\footnotesize    
$\textsuperscript{\dag}$ e-mail: \href{mailto:wangyy2027@163.com}{wangyy2027@163.com} \\    
$\textsuperscript{\ddag}$ e-mail: \href{mailto:zhaomengdie0926@163.com}{zhaomengdie0926@163.com} \\    
$\textsuperscript{\S}$ e-mail:
\href{mailto:202231140014@mail.bnu.edu.cn}{(corresponding author){xinyu.wang@mail.bnu.edu.cn}}\\   
$\textsuperscript{*}$ e-mail: \href{mailto:gpliphys@yeah.net}{(corresponding author){gpliphys@yeah.net}} 
\end{flushleft}
\maketitle

\newpage
\baselineskip 18pt
\section{Introduction}\label{sec1}
Black holes are a significant class of celestial objects predicted by general relativity. Their extremely strong gravity makes direct detection highly challenging. 
Notably, in 2015, the Laser Interferometer Gravitational-Wave Observatory (LIGO) detected gravitational waves for the first time, confirming the existence of binary stellar-mass black hole systems\cite{LIGOScientific:2016aoc}. 
In a later development, the Event Horizon Telescope (EHT) in 2019, based on very long baseline interferometry (VLBI) technology, achieved event-horizon-scale imaging of the supermassive black hole in M87 at an operating wavelength of approximately 1.3 mm\cite{EventHorizonTelescope:2019uob,EventHorizonTelescope:2022wkp}, providing the first visual evidence of the black hole.
Also, the first EHT images of the supermassive black hole at the Galactic Center, Sagittarius A* (Sgr A*), were captured in 2022\cite{EventHorizonTelescope:2022wkp,Johnson:2023ynn}.
The black hole image represents a landmark achievement, providing the first direct visual evidence of a black hole and offering an unprecedented observational tool to probe the extreme physics and strong gravity near the event horizon.

The release of black hole images has markedly advanced both theoretical and observational studies of black holes, with recent developments in shadow and imaging research offering new insights into the physical processes occurring near the event horizon.
In the EHT observations, the dark central region is called the black hole shadow, and the surrounding orange luminous ring is referred to as the bright ring. A thin and sharp ring is hidden within the region of bright ring, corresponding to the black hole’s critical curve, known as the photon ring.
The photon ring originates from light rays that orbit multiple times in the near-field region of the black hole before escaping to infinity, ultimately converging near a ring-shaped ``critical curve” on the image plane\cite{Lockhart_2022}. Evidence shows  signal is photon-ring-dominated in nearly all magnetically arrested disk models at 230 GHz on sub-Earth-diameter baselines, which has paved the way for detailed studies of the photon ring features in black holes surrounded by various astrophysical structures\cite{C_rdenas_Avenda_o_2023,Gralla_2020,Broderick:2022tfu,Shavelle:2024vwt,Tamar_2024}.
In 2019, based on a thin disk model, Wald et al. investigated the number of times light rays cross the disk and analyzed the disk image of a Schwarzschild black hole. They found that the bright ring outside black hole shadow consists of direct emission, the lensing ring, and the photon ring\cite{Gralla_2019}.
The photon ring coding the intrinsic features of spacetime, and studying it is of great significance for exploring the fundamental nature of gravity.
Under the thin-disk model, researchers have also explored the structures of lensing rings and photon rings within contexts such as loop quantum gravity, modified gravity theories, and spacetimes coupled with electromagnetic or scalar fields\cite{Xamidov:2025gcs,Zahid:2025cfu}. It has been found that the characteristics of the photon ring are closely related to parameters such as the impact parameter in loop quantum gravity and the quantum parameters in modified quantum gravity, as well as the MOG field parameters
\cite{Huang:2025gia,Zahid:2025cfu,Xamidov:2025gcs}.
For other compact objects such as boson stars, their photon rings are also influenced by the distribution of the accretion emission\cite{Rosa_2022,yang2025observational,rosa2023imaging,zhao2025light}. What is particularly important is the holographic nature of the photon ring, 
In \cite{Hadar_2022}, it is suggested that the structure of the photon ring may carry the quantum holographic information of the black hole, serving as an important bridge between classical gravity and quantum theory.
It is thus evident that, based on different accretion models, one has carefully analyzed the properties of photon ring in various black hole spacetimes, revealing their close relationship with black hole parameters and the underlying gravitational theory, thereby achieving a series of research results of significant importance.

Due to the deflection of light, black hole shadow is defined as the dark region inside the critical curve, where the light rays asymptotically approach a bound photon orbit\cite{Perlick_2022}.
The size and morphology of shadows vary significantly among different types of black holes, providing important clues for studying black hole properties\cite{chen2023black}.
For instance, the shadow size of an Einstein-Maxwell-Dilaton-Axion black hole is reduced with an increasing dilation parameter\cite{wei2013observing}; the shadow of a Kaluza-Klein rotating dilaton black hole is modulated by its mass, electric charge, and angular momentum\cite{amarilla2013shadow}; whereas the shadow of a classical Kerr black hole is also dependent on the observer's state of motion\cite{grenzebach2015aberrational}.
Moreover, one has shown that the frequencies of quasinormal modes are closely related to the photon orbits and the boundary of the black hole shadow, and that the size of the shadow can reflect the characteristics of the quasinormal modes\cite{Liu_2020,jusufi2020quasinormal}. The connection between the two provides an important physical bridge for understanding the geometric structure and oscillatory behavior of black holes.
In fact, the position of the light source has a significant impact on the black hole shadow \cite{Gralla_2019}. When the light source is located outside the critical curve, the black hole shadow coincides with the position of the critical curve. However, in realistic astrophysical environments, the accretion flow can extend close to the black hole's event horizon. In this case, the edge of the black hole shadow lies entirely within the photon ring and the critical curve, corresponding to light rays that terminate at the event horizon before crossing the equatorial plane—that is, at the location of the event horizon. This dark region is commonly referred to as the black hole’s inner shadow \cite{Chael_2021}. In 2022, Guo et al. have investigated the inner shadow of Kerr-Melvin black holes illuminated by thin accretion disks, as well as that of Kerr black holes illuminated by geometrically thick accretion disks\cite{hou2022image,zhang2024imagingthickaccretiondisks}. Additionally, researchers have conducted in-depth studies on these novel inner shadow structures in Kerr black holes under the scenario of a tilted accretion disk\cite{hu2025novel}.
It is thus evident that, based on different accretion flow models, one has conducted carefully studies of black hole accretion images and carried out in-depth analyses of their inner shadow features\cite{li2025shadow,HassanPuttasiddappa:2025tji,li2021observational,zhang2024observational,zheng2024shadow,gan2021photon,Kobialko:2025sls,chen2012properties,zeng2022shadows,li2021shadows,hou2022multi,wang2022optical,Gyulchev:2025smf,zeng2022qed,zeng2023optical,Fauzi:2024jfx,zeng2023holographic,zeng2022effects,Fauzi:2024nta,hou2024unique,zeng2023accretion,chen2024additional,Zare:2024dtf,meng2025effects,wang2024distinguishing,he2026shadow,Erices:2024lci,Chen:2025cwi,Gan:2021pwu,Pomares:2024rlj,Wang:2025dfn,Zeng:2025pch,Dyadina:2025oeg,Koam:2025tkk,Yang:2025byw,Novikov:2025elo,Olmo:2025ctf,Zeng:2025tji,Zeng:2025nmu,Alanazi:2025jof,Cai:2025rst,Xu:2025iwg,Zhang:2024hix,Luo:2025xjb,Xiong:2025hjn,Li:2025awg}, thereby laying an important foundation for more precise interpretation of inner shadow structures using EHT observational images in the future.

On the other hand, the NUT charge (or NUT parameter), often denoted as $n$, is a gravitational analogue of the magnetic monopole in electrodynamics, characterizing the gravitomagnetic mass of a source\cite{Wei_2012,Abdujabbarov_2011,Manko_2005}. 
The study of its effects is essential for deepening our understanding of fundamental aspects of gravity, such as the nature of singularities, the validity of cosmic censorship and so on \cite{Manko_2005,Johnson_2014,Hennigar_2019,Wei_2012,Duztas_2017,Zhang_2021,Chakraborty_2019,Miller:1973hqu,demianski1966combined}.
One manifestation of this effect is the corresponding black hole shadow, which is commonly considered an effective probe for revealing its significant deviations from the Kerr black hole\cite{Abdujabbarov_2012,Zhang_2021,Ghasemi_Nodehi_2021}.
For instance, according to the study of the KTN black hole shadow, it was found in 2013 that for a fixed spin parameter $a$, the presence of the NUT charge $n$ enlarges the overall size of the shadow and reduces its deformation compared with the case without $n$\cite{Abdujabbarov_2012,Zhang_2021}.
More importantly, based on the imaging data of M87*, researchers further analyzed the shadow of the KTN black hole. The results show that the current EHT observations suggest the possible existence of a NUT charge, while also ruling out the possibility that M87* is a KTN naked singularity\cite{Ghasemi_Nodehi_2021}.
Since the dark region observed in the M87* image does not correspond to the black hole shadow but to the so-called inner shadow, a detailed study of the inner shadow of KTN black holes is of great significance for probing the existence of a gravitomagnetic monopole. And, the image of KTN black hole in the context of the thin disk, particularly the inner shadow contained therein, has not been thoroughly investigated. 
Moreimportantly, the NUT charge may produce features in the inner shadow of the black hole that are more pronounced than those in the traditional black hole shadow, thereby helping to reveal the effects of a gravitomagnetic monopole and potentially allowing for their verification in future EHT observations. As a result, in this work, we adopt an accretion disk that is geometrically and optically thin and place it in the equatorial plane of black hole. Using backward ray-tracing, we perform numerical simulations of KTN black hole images, thereby obtaining thin-disk images composed of the inner shadow, photon ring, lensing ring, and direct emission. Among these features, the inner shadow stands out as the most significant indicator of the black hole’s event horizon and potential NUT charge effects.

The remaining parts of this paper are organized as follows: In Sec.\ref{sec2}, we review the KTN spacetime and discuss its inner and outer horizons. In Sec.\ref{sec23}, we provide a detailed description of the imaging method and the accretion disk model used in this study. In Sec.\ref{Sec3}, We present the images of the KTN black hole illuminated by the accretion disk and analyze the influence of the Kerr parameters and NUT parameters on the images. Finally, in Sec.\ref{sec4}, we summarize the work presented in this paper.

\section{Kerr-Taub-NUT black hole and null geodesics}\label{sec2}
In this section, we will briefly review the KTN black hole and its corresponding null geodesics. By introducing the rotating parameter \( a \) and NUT charge \( n \),  the KTN black hole is a axially symmetric type D solution,  which can be expressed in the Boyer-Lindquist coordinates\cite{chakraborty2018does}, which is
\bea
ds^2 = -\frac{\Delta}{\Sigma}(dt-\chi d\phi)^{2}+\frac{\Sigma}{\Delta}dr^2+\Sigma d\theta^2+\frac{1}{\Sigma} \sin^2 \theta \left[a dt - (r^2 + a^2 + n^2) d\phi \right]^2,
\eea 
where
\bea\label{1}
\Delta &= &r^2 - 2Mr + a^2 - n^2, \\
\Sigma &=& r^2 + (n + a \cos\theta)^2, \\
\chi &=& a \sin^2\theta - 2n \cos\theta, 
\eea
with $M$ is the black hole mass, and we have used rescaled units ($G=c=1$) through out paper. For a non-extremal KTN black hole, it's horizon is determined by $\Delta$, 
\bea
\Delta &= &r^2 - 2Mr + a^2 - n^2=0,
\eea
from which, we have
\bea
r_\pm= M\pm\sqrt{M^2 + n^2 - a^2}, 
\eea
where, $r_+$ and $r_-$ correspond to the event horizon and inner horizon of black hole.  
The true physical singularity is determined by the condition $\Sigma=0$.
When \( a^2 > n^2 \), the spacetime curvature ring singularity can be formed at $r = 0$ and $\cos{\theta} =-\frac{n}{a}$. For \( a^2 < n^2 \), there are no spacetime singularities and black hole is regular. 
For an extremal black hole, the degenerate horizon is located at \( r_{ex} = M \). If the horizons disappear \( M^2 + n^2 < a^2 \), this metric describes a naked singularity. 
In addition, when \( a = 0 \), this space time reduces to the Taub-NUT regular black hole, while the Kerr black hole is recovered by setting $n=0$.
For convenience, we denote the event horizon as \( r_h \) for use in the subsequent discussions throughout this paper. 

For the KTN spacetime, there exist two Killing vector fields $\partial_t$ and $\partial_\phi$, which correspond to two conserved quantities: the energy \( \mathcal{E} \) and the angular momentum \( \mathcal{L} \). In addition, the spacetime admits a Carter constant \( K \), and for photons, the Hamiltonian satisfies \( H = 0 \). Based on these four conditions, the equations of motion for photons can be derived as follows, 
\bea
\Sigma \dv{t}{\tau} 
    &=& \frac{\chi (\mathcal{L} - \mathcal{E} \chi)}{\sin^2\theta} + \frac{(\Sigma + a \chi)\left[(\Sigma + a \chi)\mathcal{E} - a \mathcal{L}\right]}{\Delta}, \\
    \Sigma \dv{\phi}{\tau} 
    &=& \frac{\mathcal{L} - \mathcal{E} \chi}{\sin^2\theta} + \frac{a\left[(\Sigma + a \chi)\mathcal{E} - a \mathcal{L}\right]}{\Delta}, \\
    \Sigma^2 \left( \dv{\theta}{\tau} \right)^2 
    &=& K - \frac{(\chi \mathcal{E} - \mathcal{L})^2}{\sin^2\theta} \equiv \Theta(\theta), \\
    \Sigma^2 \left( \dv{r}{\tau} \right)^2 
    &=& \left[ (\Sigma + a \chi) \mathcal{E} - a \mathcal{L} \right]^2 - \Delta K \equiv R(r).
\eea
To obtained black hole shadow, one should determine the condition of photon region where photons are filled, which are $R(r_p) = 0, R'(r_p) = 0, R''(r_p) > 0$, and $\Theta(\theta_p) = 0, \dot{\Theta}(\theta_p) = 0, \ddot{\Theta}(\theta_p) < 0$, with $r_p$ and $\theta_p$ are the radius and the latitude of the photon. With this condition, the Carter constant and  angular momentum can be energy-rescaled as \( \bar{K} = K / E^2 \) and \( \bar{L} = L_z / E \). Furthermore, by determining the celestial coordinates with respect to the reference frame and projecting them onto the Cartesian coordinates on the screen, the shadow contour of the KNT black hole can be obtained. In \cite{Zhang_2021}, the shadow structure of the KTN black hole has been investigated, and the behavior of its observables with respect to $n$ has also been carefully analyzed. Accordingly, the present work further considers the presence of a thin accretion flow extending to the event horizon along the equatorial plane of the black hole, and aims to thoroughly study the resulting inner shadow structure and observable images of the KTN black hole in such a background.

\section{Imaging method and the thin disk model}\label{sec23}

In this section, to obtain the image of the Kerr–Taub–NUT black hole, we first trace the trajectories of photons by solving their geodesic equations. For simplicity, this is accomplished through numerical integration of the geodesic equations in their Hamiltonian form. Once the photons reach the vicinity of the observer, a camera model is employed to project the incoming light rays onto the observed screen. To this end, we adopt the stereographic projection technique introduced in Ref.\cite{hou2022image}. Finally, since a black hole does not emit light by itself, an external illumination source is required to generate its image. In this work, we introduce a geometrically and optically thin accretion disk, located on the equatorial plane of the black hole, as the source of illumination for the black hole.

\subsection{Imaging method for rotating black hole}

The Hamiltonian for a photon moving along a null geodesic is given by \( H = \frac{1}{2} g_{\mu\nu} p^\mu p^\nu =0\), which implies the Hamiltonian canonical equations for photons can be expressed as,
\bea\label{p}
\dot{p}^\mu = - \frac{\partial H}{\partial x^\mu}, \quad \dot{x}^\mu = \frac{\partial H}{\partial p^\mu},
\eea
with \( x^\mu \) is the position coordinate of the photon. Obviously, we also need to determine the initial conditions for the evolution. 
For a rotating black hole, we choose a common observer to study the observed images and inner shadow of this black hole\cite{grenzebach2014photon}. When the observer located at the coordinates \((t_o, r_o, \theta_o, \phi_o)\), where a local orthonormal frame for the common observer can be constructed, which is
\bea\label{CO}
e_0&=&e_{(t)}= \frac{1}{\sqrt{ \Sigma \cdot \Delta}} \begin{pmatrix} \mathcal{A}, 0 , 0 , a \end{pmatrix}\label{CO1},\\
e_1&=&-e_{(r)}=  - \begin{pmatrix} 0 , \sqrt{\frac{\Delta}{\Sigma}} , 0 , 0 \end{pmatrix}\label{CO2},\\
e_2&=&e_{(\theta)}= \begin{pmatrix} 0 , 0 , \sqrt{\frac{1}{\Sigma}} ,0 \end{pmatrix}\label{CO3},\\
e_3&=&-e_{(\phi)}= -\frac{1}{\sqrt{\Sigma \cdot {\sin \theta}^2} } \begin{pmatrix} \chi , 0 , 0 ,1 \end{pmatrix}\label{CO4}.
\eea
In Eq.(\ref{CO}), $\mathcal{A}=r^2 + a^2 + n^2$, and the negative sign between \(e_1\)  and \(e_3\) ensures that the path of light emitted from the local orthonormal frame is reversible. In this frame, the four-momentum of a photon is expressed as
\bea
p_{(\mu)} = k_\nu e_{(\mu)}^\nu.
\eea
According to \cite{li2025shadow}, we adopt a stereographic projection method to map the photon's momentum onto an observation screen, where the stereographic projection corresponds to a camera model. For simplicity, the energy of the photon observed in the camera model is set to unity. In this model, the photon's momentum can be characterized by celestial coordinates ($\alpha$, $\beta$). Once the photon's four-momentum is fixed, the corresponding celestial coordinates are also uniquely determined, given by, 
\bea
\cos\alpha = \frac{p^{(1)}}{p^{(0)}}, \quad \tan\beta = \frac{p^{(3)}}{p^{(2)}}.
\eea
With the aid of \( (\alpha, \beta)\), the tangent vector at the the observer’s frame can be written as
\bea\label{3}
\dot{\mathcal{S}}  = \left( -{e}_0 + \cos \alpha \, {e}_1 + \sin \beta \cos \alpha \, {e}_2 + \sin \alpha \sin \beta \, {e}_3 \right),
\eea
where, the negative sign in front of \( {e}_0 \) ensures that the tangent vector points toward the past. On the other hand, the tangent vector can also be written as
\bea\label{4}
\dot{\mathcal{S}} = \dot{t} \partial_t + \dot{r} \partial_r + \dot{\theta} \partial_{\theta} + \dot{\phi} \partial_{\phi}.
\eea
By comparing Eq.\eqref{3} and Eq.\eqref{4}, one can see that when the value of a photon’s four-momentum fixed, the celestial coordinates of the image are determined. Conversely, if the celestial coordinates of the image are known, the four-momentum can be obtained through the coordinate transformation, i.e., Eqs.\eqref{CO}-\eqref{CO4}. Therefore, combining this with the position of the observer, we can give the initial values of the photon’s motion equation \( (x^\mu, p^\mu) |_{O} \).
To obtain the image of black hole, it is necessary to establish a one-to-one correspondence between the celestial coordinates \( (\alpha, \beta) \) and the points \( (x, y) \) on the screen at the observed position. In this paper, for the stereographic projection method, the standard Cartesian coordinate \( (x, y) \) at the screen can be obtained directly through simple geometric relation\cite{li2025shadow}, namely, 
\[
x_p = -2 \tan\left( \frac{\alpha}{2} \right) \sin \beta, \quad y_p = -2 \tan\left( \frac{\alpha}{2} \right) \cos \beta.
\]
To perform numerical backward ray-tracing, we also need to place pixel points on the screen and trace the light rays corresponding to each pixel.
For a camera with a field of view $\alpha_{\text{fov}}$, the length of the square screen can be expressed as, $L= 2 \tan \frac{\alpha_{\text{fov}}}{2} $. 
If the screen is divided into 
$N \times N$ pixels, the side length of each pixel is given by, $l = \frac{2}{N} \tan  \frac{\alpha_{\text{fov}}}{2}.$
For any given pixel point \( (i, j) \), its corresponding  Cartesian coordinate \( (x, y) \)  can be expressed as
\bea
x_i = l \left( \frac{i - N + 1}{2} \right), \quad y_j = l \left( \frac{j - N + 1}{2} \right),
\eea
and the celestial coordinates \( (\alpha, \beta) \) are,
\bea
&&\tan \beta= \frac{2j - (N + 1)}{2i - (N + 1)},\\
&&\tan \frac{\alpha}{2}= \frac{1}{N \tan \left( \frac{\alpha_{\text{fov}}}{2} \right)} \sqrt{\left( \frac{i - N + 1}{2} \right)^2 + \left( \frac{j - N + 1}{2} \right)^2}.
\eea
Therefore, by tracing all light rays backward through the pixels and identifying the corresponding light sources for each ray, the image of the black hole's accretion disk can be obtained.
Next, we introduce the accretion disk model in this paper, which is an optically and geometrically thin disk located at the equatorial plane of black hole, where the components of the disk can be considered as a free, electrically neutral plasma. 

\subsection{Thin disk model around black hole}\label{Sec3.2}

To describe the characteristics of the accretion flow, we first consider the motion of timelike particles.
For a timelike particle, the equation of radial motion is obtained through the normalization condition of the four-velocity \( u^\mu u_\mu = -1 \) and the conserved quantities are assumed to be \( E\) and \( L \). In this case, the radial motion can be expressed as
\bea\label{uu}
u^r=-\sqrt{\frac{V_{eff}(r,E,L)}{g_{rr}}},
\eea
where
\bea\label{veff}
V_{eff}(r)= \left( 1 + g^{tt} E^2 + g^{\phi\phi} L^2 - 2 g^{t\phi} E L \right) \bigg|_{\theta = \frac{\pi}{2}}.
\eea
In Eq.(\ref{veff}), \(V_{eff}(r)\) is defined as the effective potential, which is very important for analyzing radial motion.
For the circular orbits of timelike particles, they satisfy \( V_{eff} = \partial_r V_{eff}  =0\). considering these two conditions, the conserved quantities \( E=E_{circle}(r) \) and \( L=L_{circle}(r) \) for the circular orbit can be determined. In the following, the stability of the circular orbit is described by $\frac{d^2 V_{eff}}{dr^2} = \frac{\partial^2 V_{eff}}{\partial r^2} = V^{''}_{eff} |_{E = E_{{circle}}(r), \, L = L_{{circle}}(r)}.$
It is obvious that a stable circular orbit requires \( V^{''}_{eff} \geq 0 \), while for general asymptotically flat spacetimes, there always exists an ISCO where \(V^{''}_{eff} = 0 \). 
Therefore, for the accretion disk located on the equatorial plane, it is assumed that the accretion flow moves along circular orbits with a four-velocity outside the ISCO, where the four-velocity reads 
\bea
u^\mu_{\text{out}} &=& \left. \frac{1}{\sqrt{-g_{tt} - 2g_{t\phi} \Omega - g_{\phi\phi} \Omega^2}} \left(1, 0, 0, \Omega \right) \right|_{\theta = \pi/2},\\
\Omega &=& \frac{d\phi}{dt} = \frac{\partial_r g_{t\phi} + \sqrt{(\partial^2_r g_{t\phi}) - \partial_r g_{tt} \partial_r g_{\phi\phi}}}{\partial_r g_{\phi\phi}}.
\eea
And meanwhile, inside the ISCO, it is assumed that the accretion flow plunges inward with the specific energy and angular momentum corresponding to those at the ISCO. Those plunging trajectorie start from the ISCO and falling into the event horizon in the equatorial plane. In this sense, the radial motion equation is given by
\bea
u^r=-\sqrt{\frac{V_{eff}(r,E_{ISCO},L_{ISCO} 
)}{g_{rr}}},
\eea
where, the negative sign in the equation represents inward motion, and the value of \( r \) in the equation must satisfy \( r_H < r < r_{ISCO} \). And the correspond four-velocity reads,
\bea
u^t_{\text{in}} &=& \left. \left(-g^{tt} E_{ISCO} + g^{t\phi} L_{ISCO} \right) \right|_{\theta = \pi/2},\\
u^r_{\text{in}} &=& \left. -\sqrt{-\frac{g_{tt} u^t_{\text{in}} u^t_{\text{in}} + 2g_{t\phi} u^t_{\text{in}} u^\phi_{\text{in}} + g_{\phi\phi} u^\phi_{\text{in}} u^\phi_{\text{in}} + 1}{g_{rr}}} \right|_{\theta = \pi/2}, \\
u^\theta_{\text{in}} &=& 0, \\
u^\phi_{\text{in}} &=& \left. \left(-g^{t\phi} E_{ISCO} + g^{\phi\phi} L_{ISCO} \right) \right|_{\theta = \pi/2}.
\eea
Therefore, after specifying the accretion disk model, we proceed to discuss the accretion disk images of the KTN black hole. According to the analysis by Wald \cite{Gralla_2019}, the light ray may intersect the equatorial accretion disk once, twice, or even more than three times, corresponding to the direct image, lensing image, and higher-order images (also known as photon rings), respectively. Thus, for a rotating black hole, it is necessary to sum the radiation intensities contributed by each order of image. For simplicity, we neglect the refractive effects of the disk medium, allowing the intensity variation to be described by 
\bea\label{I1}
\frac{d}{d\lambda} \left( \frac{I_\nu}{\nu^3} \right) = j_\nu - \alpha_\nu \frac{I_\nu}{\nu^2},
\eea
where, \( I_\nu \) is the specific intensity, \( j_\nu \) is the emission coefficient, and \( \alpha_\nu \) is the absorption coefficient at frequency \( \nu \). When the photon propagate in vacuum, the absorption term \( \alpha_\nu \) and the emission term \( j_\nu \) are all equal to zero, \( \frac{I_\nu}{\nu^3} \) is a conserved quantity along the photon geodesic. 
We assume that the accretion disk is stable, axisymmetric, and has $Z_2$
symmetry about the equatorial plane. Additionally, since the accretion disk is geometrically and optically thin, when light rays pass through the disk and absorption effects can be neglected, Eq.(\ref{I1}) simplifies to
\bea\label{Ino}
I_{\nu_0} = \sum_{m=1}^{M_{\text{max}}} f_m g_m^3 J_m.
\eea  
In Eq.(\ref{Ino}), the symbol $m$ represents the number of intersections between the light ray and the accretion disk, $M_{max}$ denotes the maximum number of intersections, $f_m$ are the fudge factors, which is fixed as 1 in this paper. 
Based on the observed images' datas of M87\text{*} and Sgr A\text{*}, here we employ a specific form of emissivity $J$,
\bea
J = \exp\left( -\frac{1}{2} \left(\log\frac{r}{r_h}\right)^2 - 2 \log\frac{r}{r_h}
\right).
\eea
When \( r \geq r_{\text{ISCO}} \), the redshift factor can be expressed as
\bea
g_{out} = \frac{e}{\zeta (1 - \Omega_m \cdot {\mathcal{L}}/ {\mathcal{E}})} , \quad r \geq r_{\text{ISCO}}.
\eea
When \( r < r_{\text{ISCO}} \), we have
\bea
g_{in} = -\frac{e}{u^r_{\mathfrak{in}} p_r / \mathcal{E} + E_{\text{ISCO}}(g^{tt} - g^{t\phi} \cdot {\mathcal{L}}/ {\mathcal{E}}) + L_{\text{ISCO}}(g^{\phi\phi} \cdot {\mathcal{L}}/ {\mathcal{E}} - g^{t\phi})}, \quad r < r_{\text{ISCO}},
\eea
where, $\Omega_m = (u_{in}^{\phi}/u_{in}^t)|_{r=r_m}$, $\zeta = \left. \sqrt{\frac{-1}{g_{tt} + 2g_{t\phi}\Omega_m + g_{\phi\phi}\Omega_m^2}} \right|_{r=r_m}$, and for asymptotically flat spacetimes $e=1$, \( u^t \), \( g_{tt} \), \( g_{t\phi} \), and \( g_{\phi\phi} \) take the value at \( r = r_m \).

With the aid of the accretion disk model and imaging method described above, we can numerically solve the geodesic equations to generate thin-disk images of the KTN black hole. This allows us to extract the influence of the NUT charge on the black hole shadow and image structure, analyze its observable spacetime features, and provide a possible reference for testing the existence of the NUT charge through astronomical observations.

\section{Images of KTN black hole under the thin disk}\label{Sec3}

\begin{figure}[H]
    \centering
    \begin{subfigure}[b]{0.3\textwidth}
        \includegraphics[width=\textwidth]{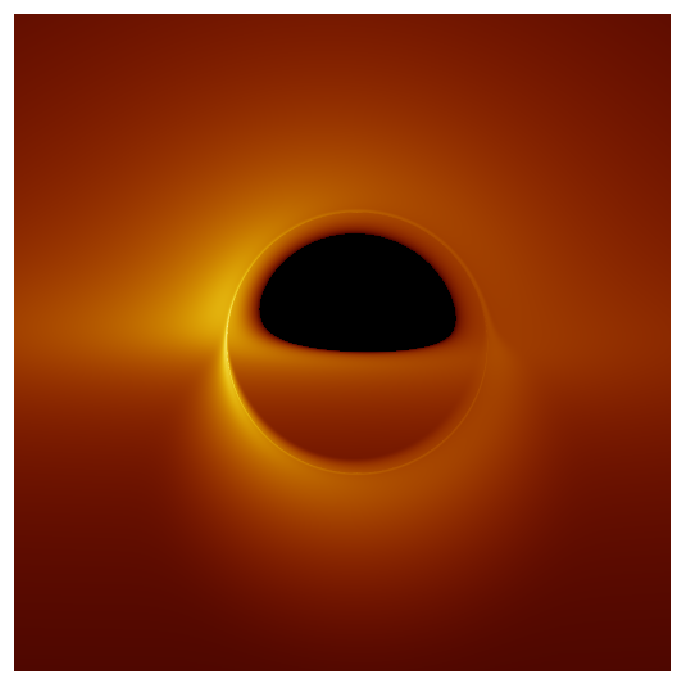}
        \caption{$a$=0.2, $n$=0.1}
        \label{4a}
    \end{subfigure}
    \hfill
    \begin{subfigure}[b]{0.3\textwidth}
        \includegraphics[width=\textwidth]{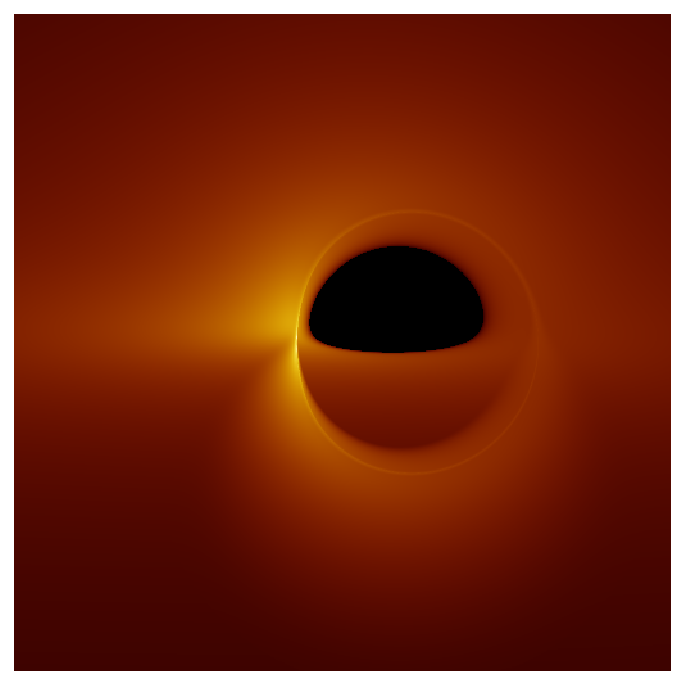}
        \caption{$a$=0.94, $n$=0.1}
        \label{4b}
    \end{subfigure}
    \hfill
    \begin{subfigure}[b]{0.3\textwidth}
        \includegraphics[width=\textwidth]{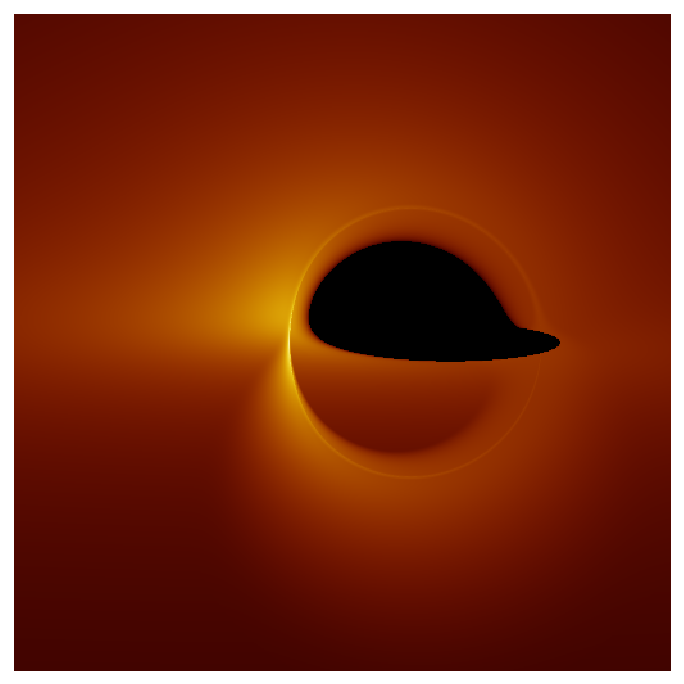}
        \caption{$a$=0.94, $n$=0.3}
        \label{4c}
    \end{subfigure}

    \begin{subfigure}[b]{0.3\textwidth}
        \includegraphics[width=\textwidth]{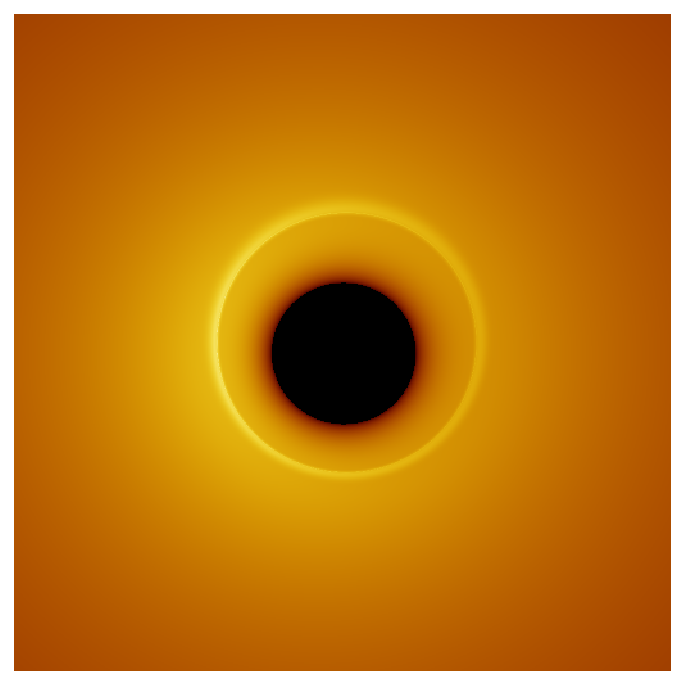}
        \caption{$a$=0.2, $n$=0.1}
        \label{4d}
    \end{subfigure}
    \hfill
    \begin{subfigure}[b]{0.3\textwidth}
        \includegraphics[width=\textwidth]{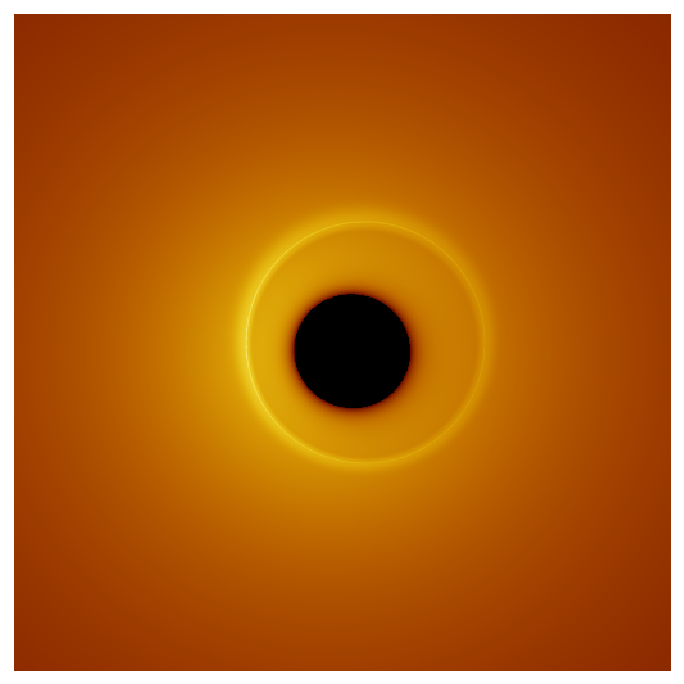}
        \caption{$a$=0.94, $n$=0.1}
        \label{4e}
    \end{subfigure}
    \hfill
    \begin{subfigure}[b]{0.3\textwidth}
        \includegraphics[width=\textwidth]{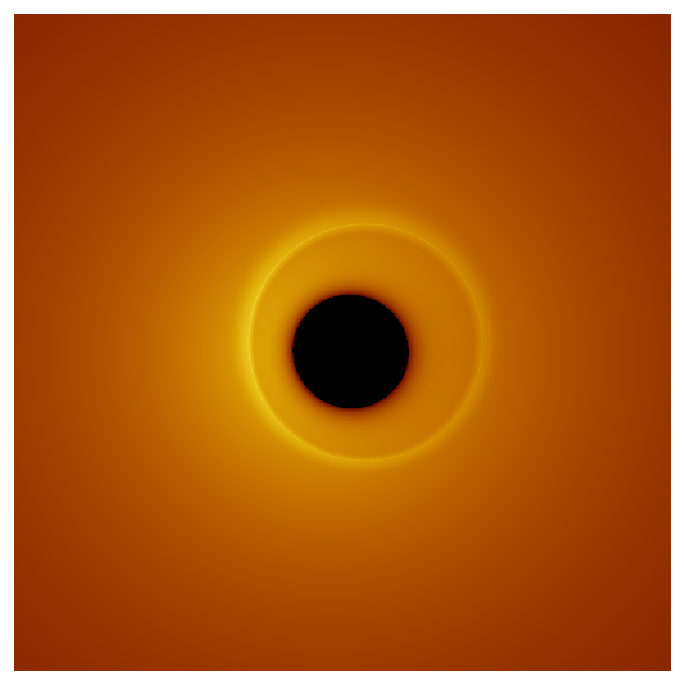}
        \caption{$a$=0.94, $n$=0.3}
        \label{4f}
    \end{subfigure}
    \caption{Images of KTN black hole with the upper and lower rows correspond to $\theta_o = 80^\circ$ and $\theta_o = 163^\circ$, respectively.}
    \label{4o}
\end{figure}

In this section, we numerically simulate the thin-disk images of a KTN black hole using the backward ray-tracing technique, following the imaging methodology and disk model described in Sec.\ref{Sec3.2}. The observer is positioned at $r = 500$ with the angles $\theta_o = 80^\circ$ and $163^\circ$, which will be used for later analysis.
Given the rapid radial decay of the accretion disk’s emissivity profile, the outer boundary is set at $r = 20$, while the event horizon serves as the inner boundary. The image resolution is set to $512 \times 512$, which is sufficient to capture the distinctive imaging features of KTN black holes. With those considerations, the images of KTN black hole under the prograde accretion flow are presented in Fig.\ref{4o}, they are

In Fig.\ref{4o}, the completely dark region at the center of each image results from the direct image of event horizon of black hole, commonly referred to as the ``inner shadow". And, the bright, narrow curves surrounding this region are known as  the ``photon ring", formed due to the strong gravitational lensing effects originated from black hole. 
It is obvious that at $\theta_o = 80^\circ$, the direct images and lensed images can be clearly distinguished, whereas at $\theta_o = 163^\circ$, they become difficult to distinguish. Nevertheless, in both cases ($\theta_o = 80^\circ$ and $163^\circ$), the inner shadow and photon ring remain clearly observable. 
In all subfigures of Fig.\ref{4o}, the left side of the photon ring appears significantly brighter than the right side. This brightness asymmetry is a result of the Doppler effect of the accretion disk - the image of the accretion flow on the side approaching the observer appears brighter, while that on the side moving away from the observer appears dimmer.
For the Fig.\ref{4a} and Fig.\ref{4b} with fixed NUT parameter, one can see that the critical curve (also referred to as the ``photon ring") on the left side of Fig.\ref{4b} appears flatter compared to Fig.\ref{4a}, and this flattening leads to a progressive deformation toward a ``D"-shaped profile.
In addition, as the spin parameter $a$ increases, the inner shadow gradually transforms from a regular semicircle shape into a right-tilted semicircle, while its overall size progressively decreases.
Also, by comparing Fig.\ref{4c} with Fig.\ref{4b}, we can see that the case with $n=0.3$ shows a larger inner shadow and critical curve than the case $n=0.1$. 
Even more interestingly, it is obvious that Fig.\ref{4c} shows greater brightness on the left side compared to Fig.\ref{4b}, and its inner shadow displays more pronounced outward distortion for $n=0.3$. Compared to the standard semicircle inner shadow, the lower-right edge of this structure is significantly stretched and sharply protrudes outward, causing the black region to extend beyond the critical curve and take on a shape resembling a right-tilted ``duck-cap–like" shape. 
This morphological feature represents a newly identified type of inner shadow region, which we refer to as the “extended inner shadow” of black hole, which is not previously observed.
In the second row of images, the observer is positioned relatively close to the south pole, so the inner shadow appears circular even when the parameter $n = 0.3$.
For Fig.\ref{4d} and Fig.\ref{4e} with fixed $n=0.1$, the inner shadow and critical curve for $a=0.94$ are smaller than for the case $a=0.2$. Similarly, comparing Fig.\ref{4e} and Fig.\ref{4f}, the case with $n=0.3$ shows larger inner shadow and critical curves than $n=0.1$. This implies that the inner shadow and critical curve of KTN black hole decrease with $a$ increases, while increase with the NUT parameter $n$. 
Then, in the following, we also present the corresponding redshift images of Fig.\ref{4o}, together with its the corresponding maximum value of blueshift of direct images, as $\theta_o=80^\circ$.

\begin{figure}[H]\label{f2}
    \centering
    \begin{subfigure}[b]{0.3\textwidth}
        \includegraphics[width=\textwidth]{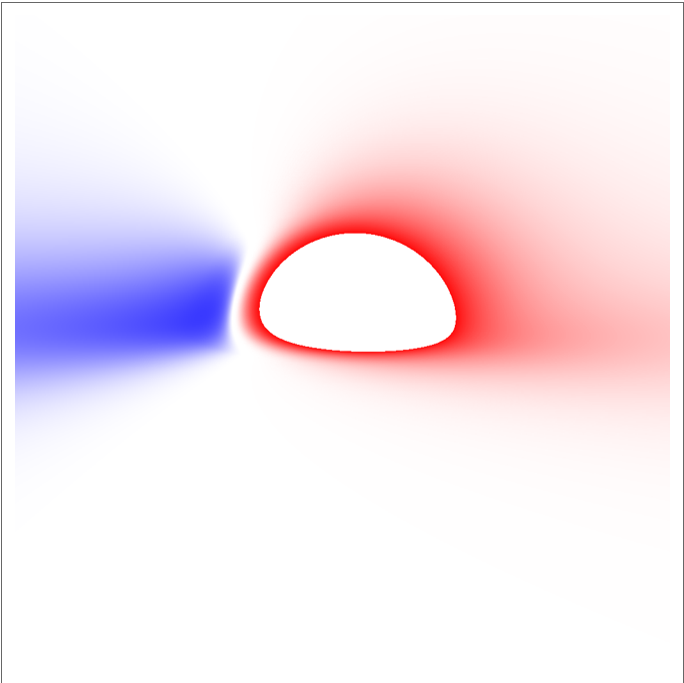}
        \caption{$a$=0.2, $n$=0.1}
        \label{6a}
    \end{subfigure}
    \hfill
    \begin{subfigure}[b]{0.3\textwidth}
        \includegraphics[width=\textwidth]{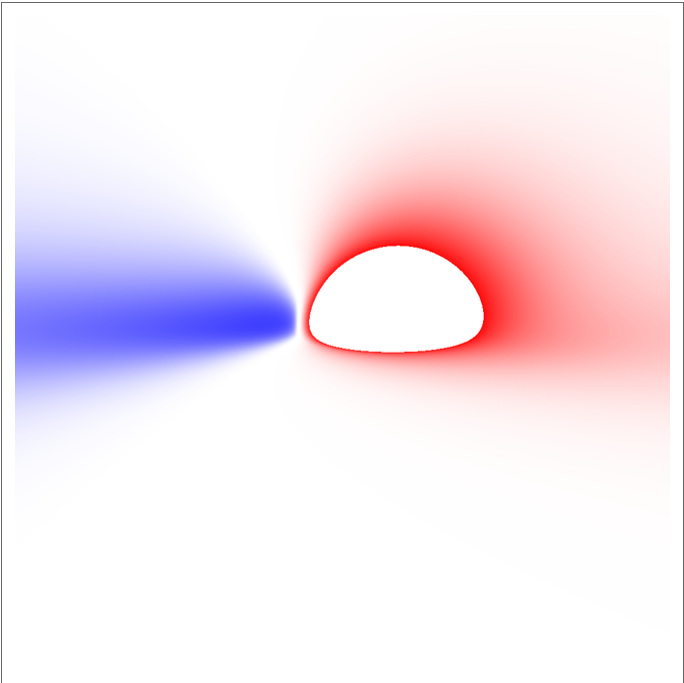}
        \caption{$a$=0.94, $n$=0.1}
        \label{6b}
    \end{subfigure}
    \hfill
    \begin{subfigure}[b]{0.3\textwidth}
        \includegraphics[width=\textwidth]{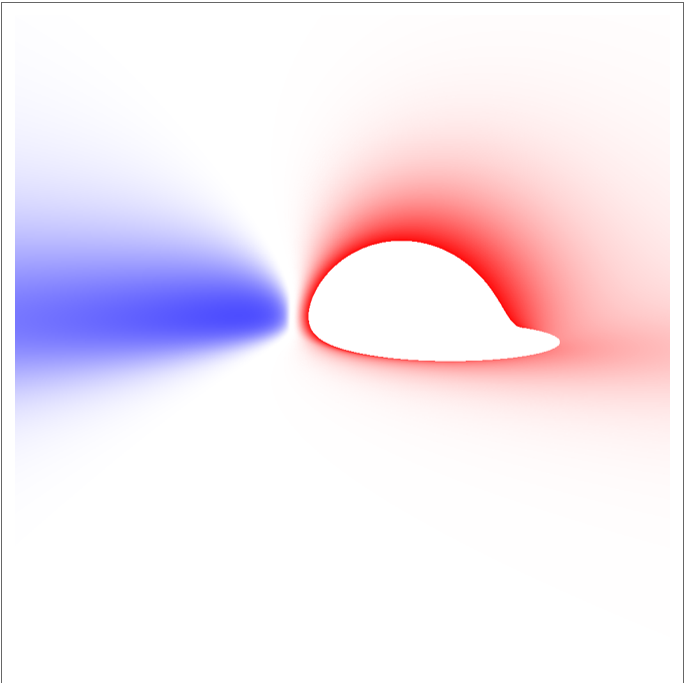}
        \caption{$a$=0.94, $n$=0.3 }
        \label{6c}
    \end{subfigure}
\begin{subfigure}[b]{0.3\textwidth}
        \includegraphics[width=\textwidth]{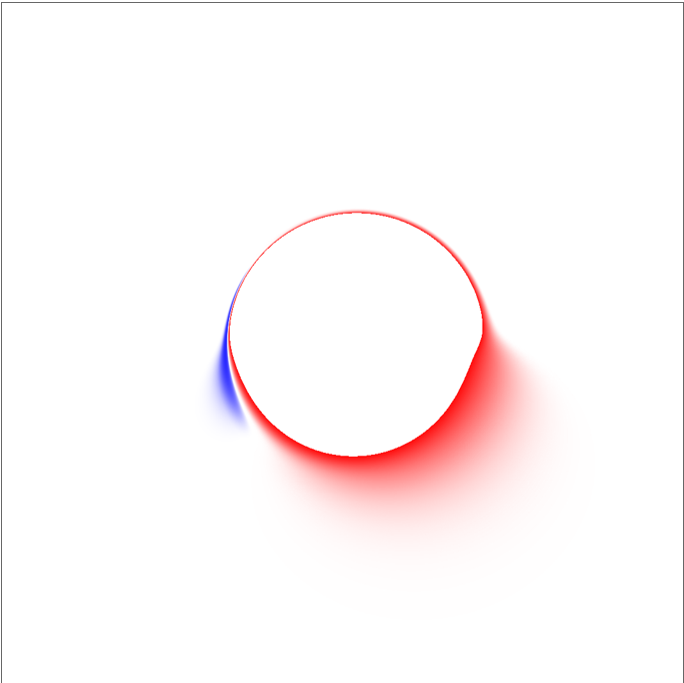}
        \caption{$a$=0.2, $n$=0.1}
    \end{subfigure}
    \hfill
    \begin{subfigure}[b]{0.3\textwidth}
        \includegraphics[width=\textwidth]{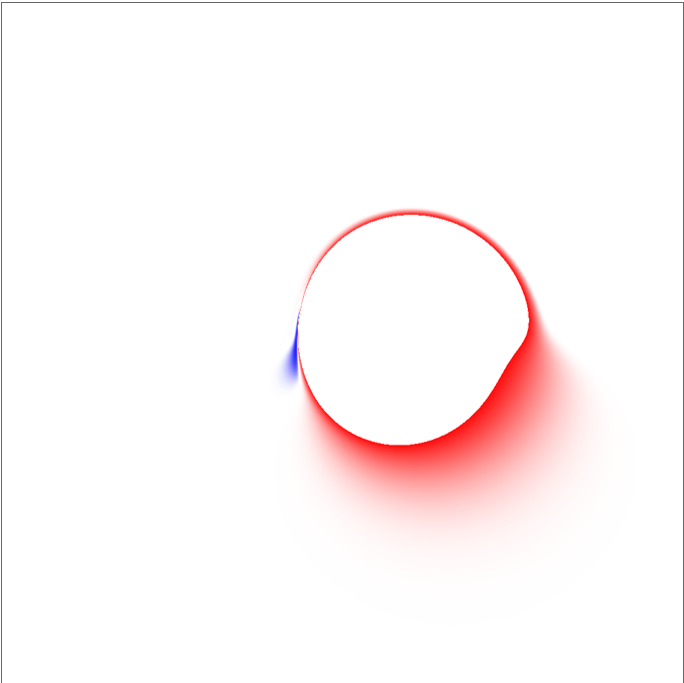}
        \caption{$a$=0.94, $n$=0.1}
    \end{subfigure}
    \hfill
    \begin{subfigure}[b]{0.3\textwidth}
        \includegraphics[width=\textwidth]{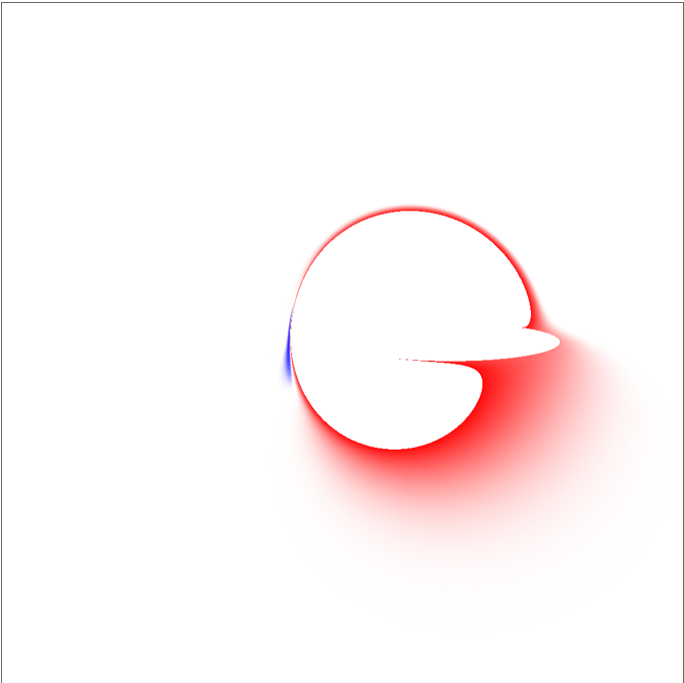}
        \caption{$a$=0.94, $n$=0.3}
    \end{subfigure}
    \caption{The redshift factors of the direct (the first row) and Lensed (the second row)images from the KTN black hole's accretion disk, where red and blue colors represent redshift and blueshift, respectively.}
    \label{6r}
\end{figure}

\begin{table}[H]
\centering
\caption{Maximum blueshift ($g_{\max}$) of direct images for different spin and NUT parameters. Observation angle is $\theta_o = 80^\circ$.}
\label{tab:redshift}
\begin{tabular}{cccccc}
\toprule
\toprule
\multicolumn{2}{c}{Parameter} & \multicolumn{1}{c}{$a = 0.2$} & \multicolumn{2}{c}{$a = 0.94$} \\
\cmidrule(r){1-2} \cmidrule(lr){3-3} \cmidrule(l){4-5}
$n$ &  & 0.1 & 0.1 & 0.3 \\
\midrule
\multirow{2}{*}{$g_{\max}$} & direct & 1.41341 & 1.39028 & 1.33826 \\
 & lensed & 1.54730 & 1.60282 & 1.52530 \\
\addlinespace
\bottomrule
\bottomrule
\end{tabular}\label{t1}
\end{table}

In Fig.\ref{6r}, the white area enclosed by the red region corresponds to the inner shadow in both the direct and lensed images, with a dark red region closely surrounding it. 
From subfigures (a) and (b) of Fig.\ref{6r}, it can be seen that as the spin parameter $a$ increases, the redshift region surrounding the inner shadow gradually decreases. From subfigures (b) and (c), it is evident that with the increase of the NUT parameter $n$, the shape of the inner shadow undergoes a significant deformation, which in turn directly modifies the redshift region. From subfigures (d), (e), and (f), one can observe that as both $a$ and $n$ increase, the blueshift region shrinks noticeably, and the morphology of the inner shadow in the corresponding lensed images also exhibits a pronounced variation.
Furthermore, Tab.\ref{t1} presents the maximum blueshift of the direct images for an observer at $80^\circ$, showing that both the spin parameter $a$ and the NUT parameter $n$ reduce the maximum blueshift.
In a word, it can be seen that both the spin parameter and the nut parameter significantly influence the redshift factor, thereby affecting the accretion disk image of the black hole.
In addition, we further analyze the variation of the observable intensity of the thin disk along the $X$ and $Y$ axes for an observer at $\theta_o = 80^\circ$, so that illustrate the impact of $n$ more clearly. 

\begin{figure}[H]
    \centering
    \begin{subfigure}[b]{0.3\textwidth}
        \includegraphics[width=\textwidth]{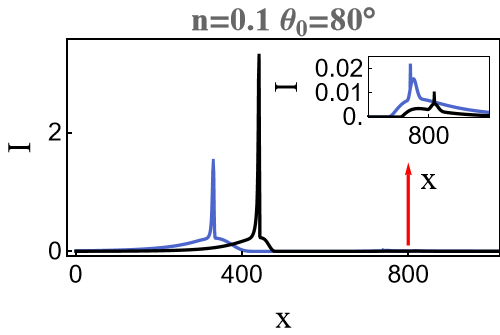}
        \caption{}
        \label{8a}
    \end{subfigure}
    \hfill 
    \begin{subfigure}[b]{0.3\textwidth}
        \includegraphics[width=\textwidth]{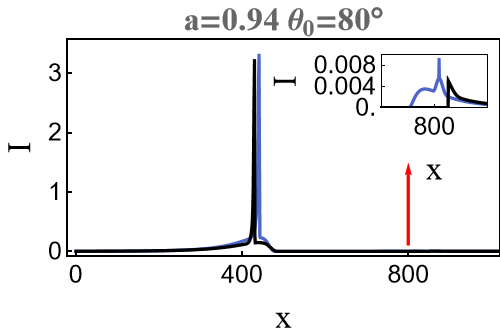}
        \caption{}
        \label{8b}
    \end{subfigure}
    \hfill
    \begin{subfigure}[b]{0.3\textwidth}
        \includegraphics[width=\textwidth]{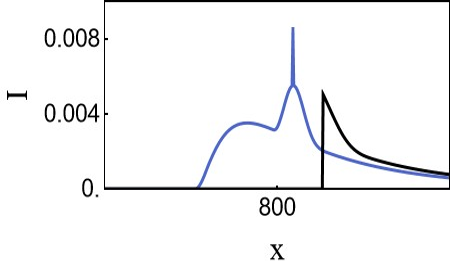}
        \caption{}
        \label{8c}
    \end{subfigure}
    \caption{The intensity distribution along the x-axis on the screen.}
    \label{8i}
\end{figure}

\begin{figure}[H]
    \centering
    \begin{subfigure}[b]{0.48\textwidth}
        \includegraphics[width=\textwidth]{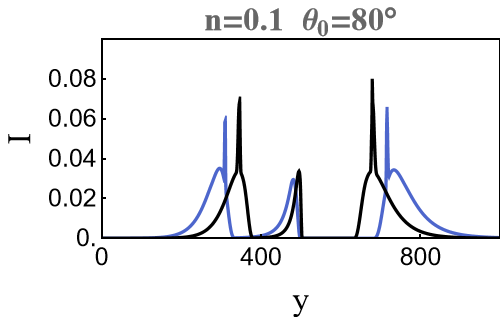}
        \caption{}
        \label{9a}
    \end{subfigure}
    \hfill 
    \begin{subfigure}[b]{0.48\textwidth}
        \includegraphics[width=\textwidth]{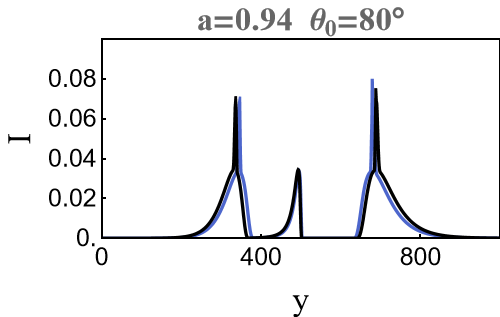}
        \caption{}
        \label{9b}
    \end{subfigure}
    \caption{The intensity distribution along the y-axis on the screen.}
    \label{9i}
\end{figure}

In subfigure (a) of Fig.\ref{8i}, the blue curve represents $a=0.2$ and the black curve $a=0.94$, while in subfigure (b), the blue and black lines correspond to $n=0.1$ and $n=0.3$, respectively.
As the spin parameter $a$ increases, the intensity distribution along the X-axis shows that both the left and right observed peaks shift to the right, with the left peak exhibiting a greater displacement.
For the NUT parameter $n$, an increase in $n$ appears to widen the separation between the two observed intensity peaks. Notably, for $n = 0.3$, the right peak initially rises sharply and then gradually decreases, which differs from the behavior observed for $n = 0.1$, where the magnified details in the inset of subfigure(b) are presented in subfigure(c). This phenomenon is likely attributable to changes in the black hole's inner shadow. And, the intensity distribution observed along the \(Y\)-axis exhibits only minor variations. As the spin parameter \(a\) decreases, the separation between the intensity peaks becomes smaller, whereas an increase in the NUT parameter \(n\) leads to a slight increase in the peak separation.
We can see that the NUT parameter $n$ modulates the peak separation and influences the local behaviour of the right peak-effects that are particularly pronounced along the $X$-axis and relatively minor along the $Y$-axis. These features result in observed images that differ significantly from those of a Kerr black hole. 

\begin{figure}[H]
    \centering
    \begin{subfigure}[b]{0.3\textwidth}
        \includegraphics[width=\textwidth]{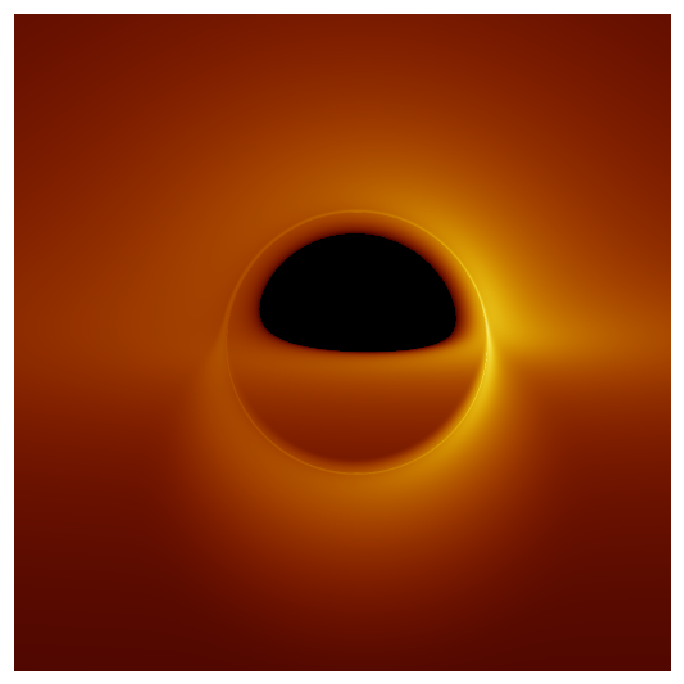}
        \caption{$a$=0.2 $n$=0.1}
        \label{fig:image1}
    \end{subfigure}
    \hfill 
    \begin{subfigure}[b]{0.3\textwidth}
        \includegraphics[width=\textwidth]{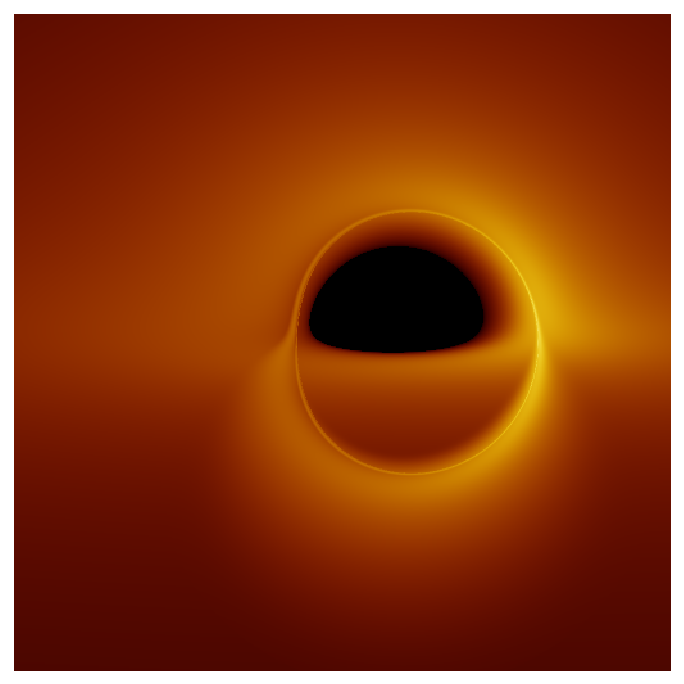}
        \caption{$a$=0.94 $n$=0.1}
        \label{fig:image2}
    \end{subfigure}
    \hfill
    \begin{subfigure}[b]{0.3\textwidth}
        \includegraphics[width=\textwidth]{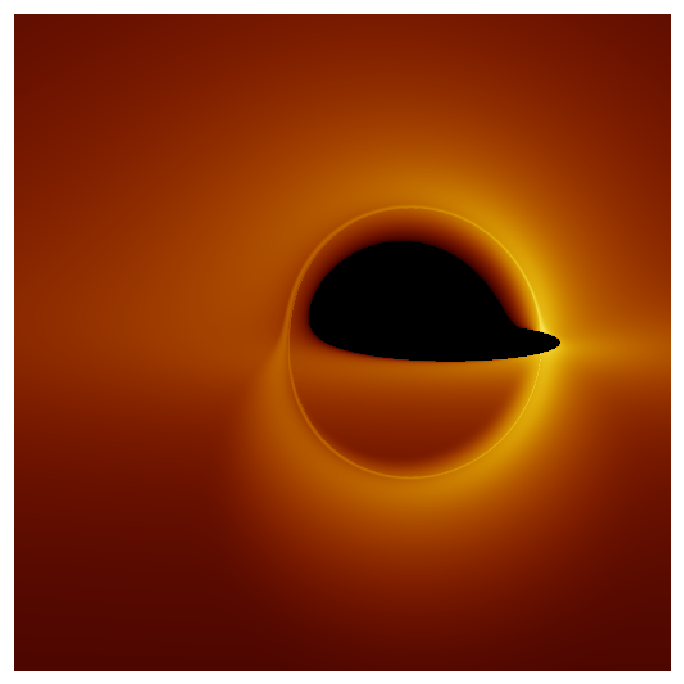}
        \caption{$a$=0.94 $n$=0.3}
        \label{fig:image3}
    \end{subfigure}
    \caption{The images of a KTN black hole illuminated by a retrograde accretion flow, corresponding to $\theta_o = 80^\circ$.}
    \label{5i}
\end{figure}

For the retrograde flow model, the resulting observable image of the KTN black hole is also shown in Fig.\ref{5i}.
We observe that in retrograde flows, the variation patterns of the inner shadow and critical curve with respect to the Kerr parameter $a$ and NUT parameter $n$ are completely consistent with those in prograde flows. However, the key difference lies in the brightness distribution: prograde flows exhibit significantly higher intensity on the left side of the disk, while due to the Doppler effect, retrograde flows show the opposite energy concentration characteristic, with the right side displaying markedly higher intensity than the left. Close examination reveals that the brightness distribution in retrograde disks is more dispersed on both sides, whereas prograde disks show more concentrated brightness on the left side. Additionally, the brightness on the right side of retrograde disks is lower than that on the left side of prograde disks. Consequently, prograde flows present strong left-side contrast with sharp edges, while retrograde disks exhibit less dramatic contrast and relatively smoother transitions, which implies that the thin-disk images of the KTN black hole not only faithfully reflect the intrinsic features of the spacetime, but also effectively reveal the properties of the accretion disk itself.

\begin{figure}[H]
    \centering
    \begin{subfigure}[b]{0.3\textwidth}
 \includegraphics[width=\textwidth]{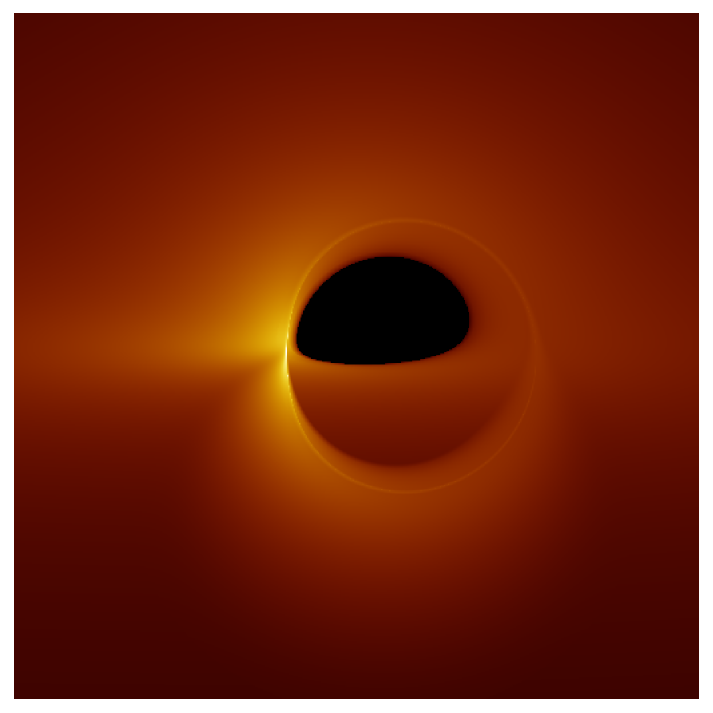}
        \caption*{$a$=0.94 $n$=0 (Kerr)}
    \end{subfigure}
    \hfill
    \begin{subfigure}[b]{0.3\textwidth}
        \includegraphics[width=\textwidth]{Fig6/a=0.94n=0.180.png}
        \caption*{$a$=0.94 $n$=0.1}
    \end{subfigure}
    \hfill
    \begin{subfigure}[b]{0.3\textwidth}
        \includegraphics[width=\textwidth]{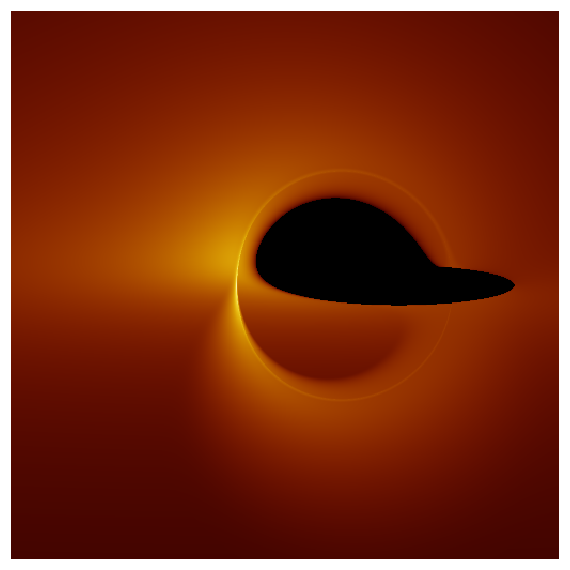}
        \caption*{$a$=0.94 $n$=0.4}
    \end{subfigure}
    \caption{Images of Kerr and KTN black holes illuminated by a prograde flows for $\theta_o = 80^\circ$.}
    \label{10o}
\end{figure}

\begin{figure}[H]
    \centering
    \begin{subfigure}[b]{0.48\textwidth}
        \includegraphics[width=\textwidth]{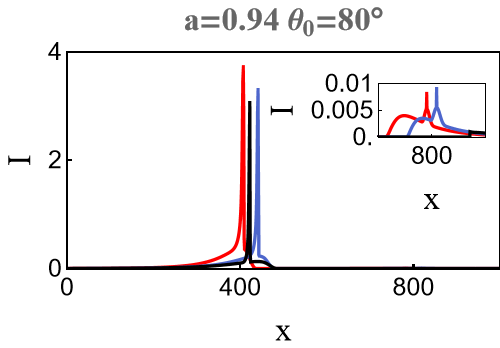}
        \caption{}
        \label{11a}
    \end{subfigure}
    \hfill 
    \begin{subfigure}[b]{0.48\textwidth}
        \includegraphics[width=\textwidth]{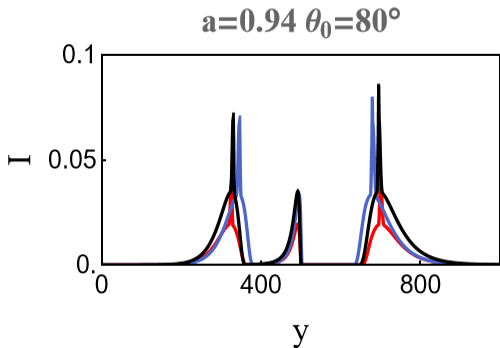}
        \caption{}
        \label{11b}
    \end{subfigure}
    \caption{The intensity distribution of Kerr and KTN black hole along the x-axis and y-axis.}
    \label{11i}
\end{figure}

Finally, we note that the results shown in Fig.\ref{4o} indicate that the inner shadow of the KTN black hole undergoes significant variations, taking on a right-tilted, ``duck-cap–like" shape. To examine the structural features of this inner shadow in greater detail, we performed a comparative analysis of the thin-disk images of the KTN and Kerr black holes, with the results presented in Figs.\ref{10o} and \ref{11i}.

In Fig.\ref{11i}, the red curve corresponds to the Kerr black hole ($n=0$), the blue curve to $n=0.1$, and the black curve to $n=0.4$.
One can see that the influence of the NUT parameter on the intensity distributions are completely consistent with those obtained in Fig.\ref{9i}. 
And, the results from Fig.\ref{10o} show that the inner shadow of Kerr black hole is, overall, approximately semicircular, with a well-defined boundary and good symmetry, consistent with its axisymmetric nature. The rotating effect may cause the shadow to shift slightly toward the direction of left side, but the overall morphology still essentially retains a semicircular structure. 
When $n = 0.1$, the inner shadow gradually evolves from a semicircular shape shifted in the direction of left side to an almost regular semicircular structure.
More importantly, as $n$ further increases to $0.4$, the critical curve gradually evolves into a regular circle, while the inner shadow undergoes significant changes, exhibiting a right-tilted, ``duck-cap-like" shape, which is completely distinct from the inner shadows of Kerr and other black holes \cite{he2026shadow,hou2022image,Zeng:2025pch}. 
We refer to this characteristic inner shadow structure as the ``extended inner shadow" of black hole\cite{chael2021observing}. 
Clearly, according to definitions in the original literature, the inner shadow of a black hole generally represents the size of the event horizon. However, in this work, we find that the dark region in the thin-disk image of a KTN black hole does not fully correspond to the black hole's event horizon. Therefore, we refer to it as the ``extended inner shadow" of the KTN black hole. Specifically, part of the extended inner shadow-namely, the region corresponding to the tongue of the right-tilted ``duck-cap–like" shape-does not represent the event horizon. This arises because the photons corresponding to this region on the observer's screen do not cross the equatorial plane. Meanwhile, the accretion disk's light source is located on the equatorial plane, so the light reaching the observer from this region does not receive contributions from the light source. In contrast, for a Kerr black hole, photons in the corresponding region can cross the equatorial plane, so this region exhibits observable light intensity in the Kerr black hole image. To gain a clearer understanding of the formation mechanism of the extended inner shadow, several representative pixels were selected within the inner shadow region. Their trajectories were then traced in the \(x\!-\!z\) plane, and the results are presented in the figure below.

\begin{figure}[H]
    \centering
    \begin{subfigure}[b]{0.45\textwidth}
        \includegraphics[width=\textwidth]{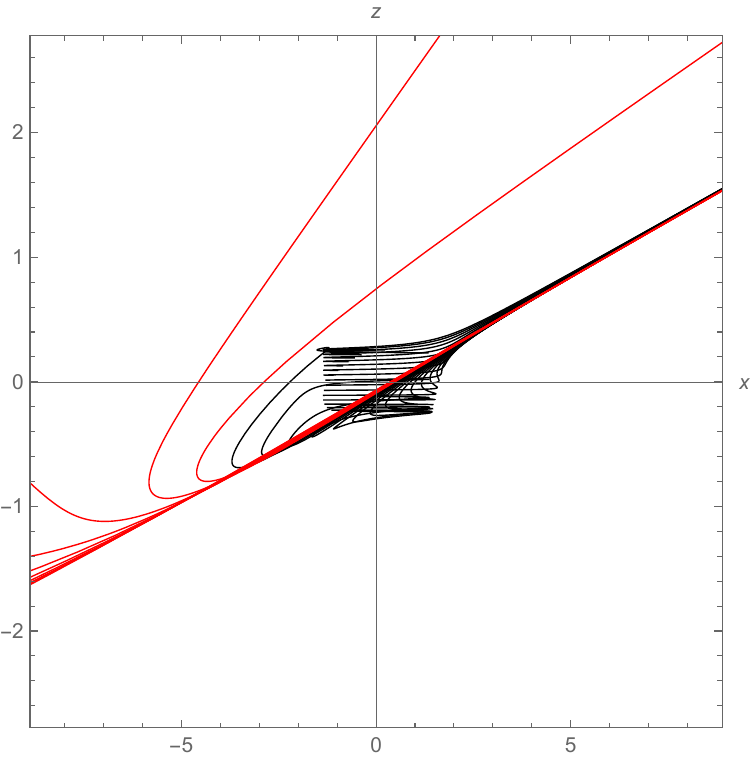}
        \caption{$a$=0.94 $n$=0}
        \label{12a}
    \end{subfigure}
    \hfill 
    \begin{subfigure}[b]{0.45\textwidth}
        \includegraphics[width=\textwidth]{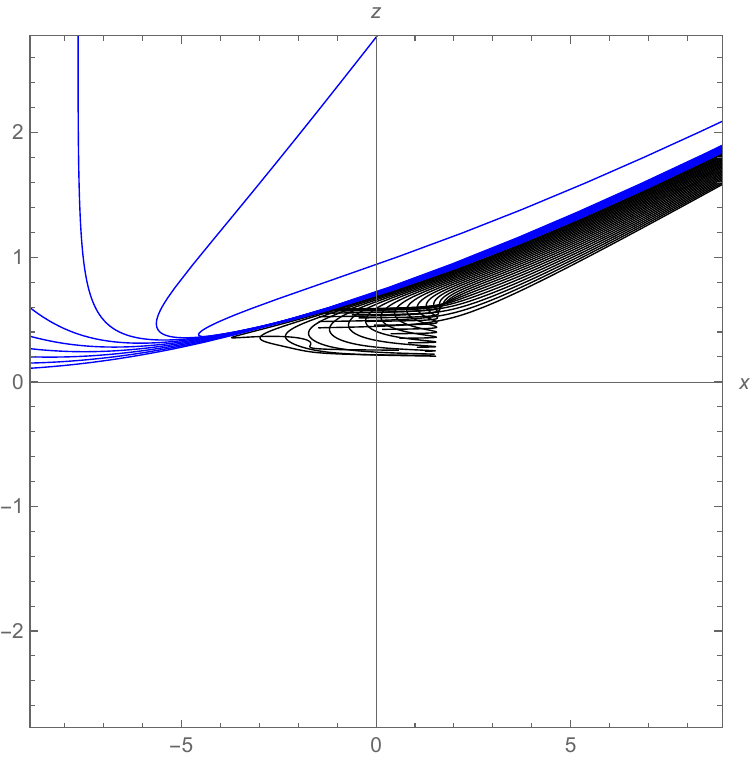}
        \caption{$a$=0.94 $n$=0.4}
        \label{12b}
    \end{subfigure}
    \caption{The light trajectories along the \(x\)-axis inside the extended inner shadow region.}
    \label{12i}
\end{figure}

\begin{figure}[H]
    \centering
    \includegraphics[width=0.4\textwidth]{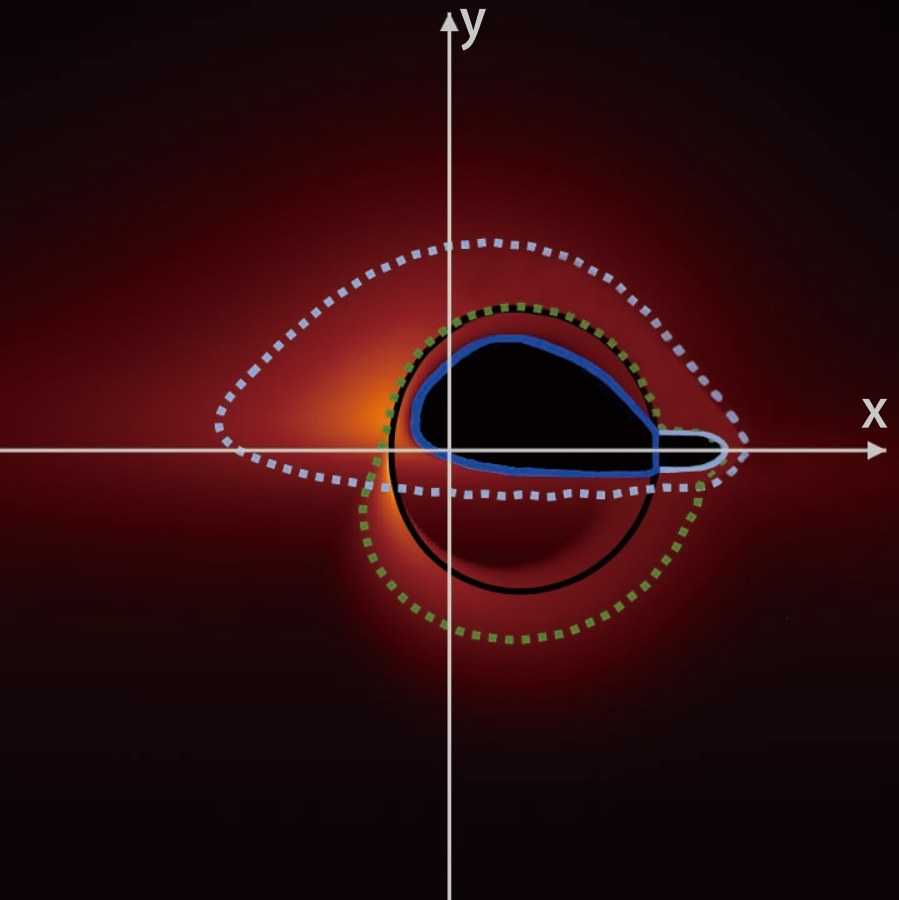} 
    \caption{An illustration of the image of the KTN black hole illuminated for $\theta_o = 80^\circ$, $a = 0.94$ and $n = 0.4$.}
    \label{fig:my_label} 
\end{figure}

In Fig.\ref{12i}, the red curves represent the light rays that cross the equatorial plane, the blue curves denote those that do not cross the equatorial plane, and the black curves correspond to the light rays that fall into the black hole. 
The scenario depicted in subfigure(a) a actually corresponds to the Kerr black hole.
The subfigure(a)  the Kerr black hole, the light trajectories consist solely of red and black curves, whereas subfigure(b) illustrates the case of the KTN black hole.
It can be observed that, for the same set of pixels, the corresponding light rays in the Kerr black hole are either absorbed by the black hole or cross the equatorial plane; whereas in the KTN black hole, the situation differs: some rays are still absorbed by the black hole, while others no longer cross the equatorial plane. This indicates that light rays which would have crossed the equatorial plane in the Kerr black hole do not do so in the KTN black hole. This is precisely the reason for the formation of the ``extended inner shadow".
Based on the above analysis, it can be seen that the newly emerging ``extended inner shadow" in the KTN black hole exhibits significant differences from the inner shadow of the Kerr black hole. This observable effect makes it possible to directly distinguish between Kerr and KTN black holes, thereby providing a new avenue for assessing the possible existence of the NUT charge in the universe. 
To more clearly illustrate the morphological features of the extended inner shadow in the thin-disk image of the KTN black hole, we distinguish the black hole horizon from the extended inner shadow region in the schematic diagram shown in Fig.\ref{fig:my_label}, and label the direct image, lensed image, and critical curve.

\begin{figure}[!h]
\centering
\setlength{\tabcolsep}{-0.2pt}
\setlength{\fboxsep}{0.5pt}
\renewcommand{\arraystretch}{-0.5}
\begin{tabular}{c c c c}
      & $n=0$ & $n=0.1$ & $n=0.3$ \\ \\ & \textcolor{white}{..} & \textcolor{white}{..} & \textcolor{white}{..}\\ \\ & \textcolor{white}{..} & \textcolor{white}{..} & \textcolor{white}{..}\\
      \\ & \textcolor{white}{..} & \textcolor{white}{..} & \textcolor{white}{..}\\
      
    \raisebox{5\height}&
    \fbox{\includegraphics[width=0.23\textwidth]{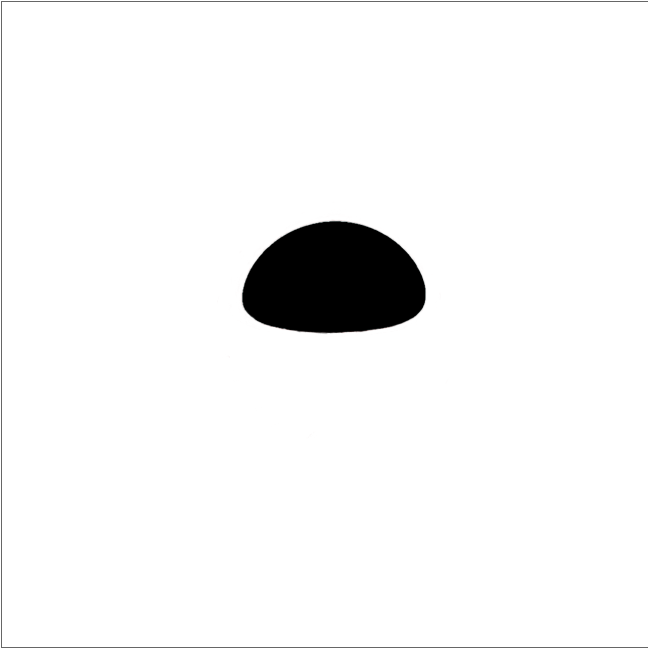}} &
    \fbox{\includegraphics[width=0.23\textwidth]{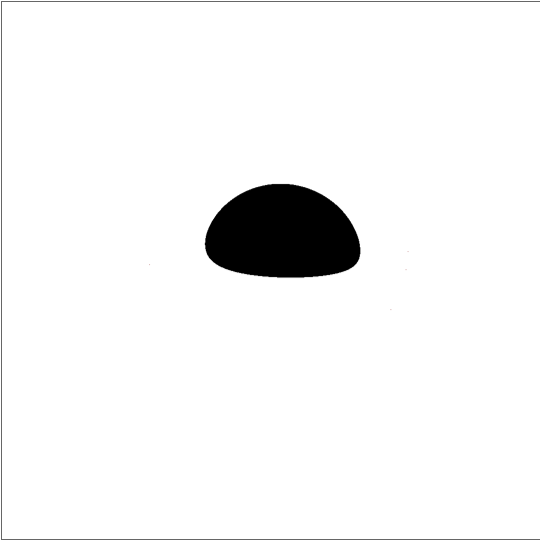}} &
   \fbox{\includegraphics[width=0.23\textwidth]{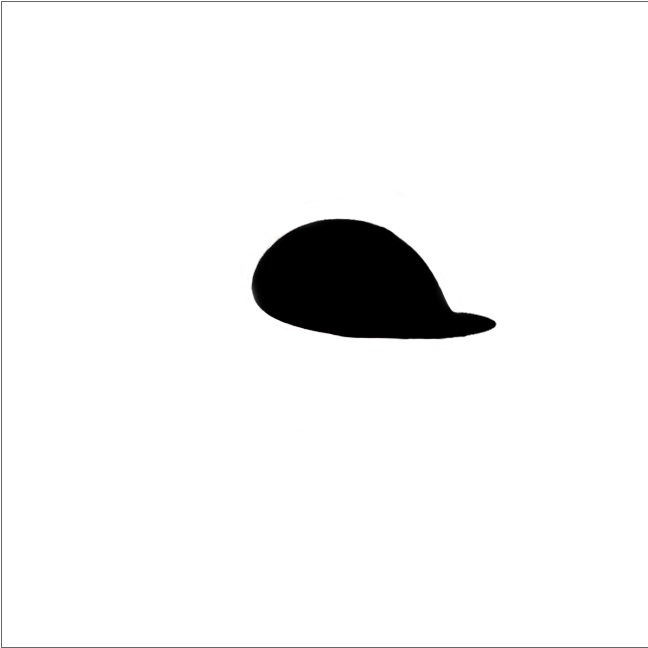}} \\
    \raisebox{5\height} &
    \fbox{\includegraphics[width=0.23\textwidth]{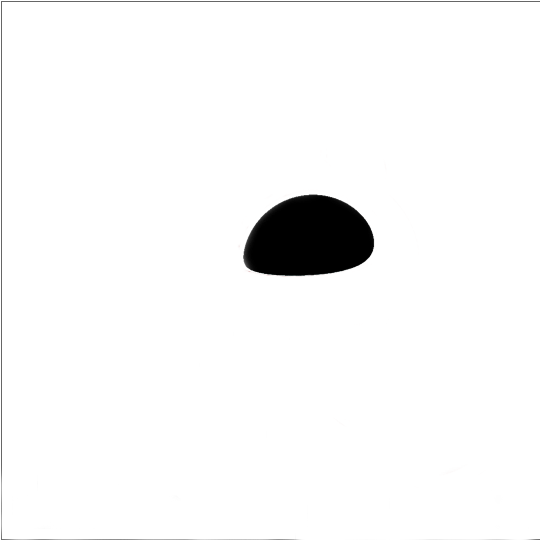}} &
    \fbox{\includegraphics[width=0.23\textwidth]{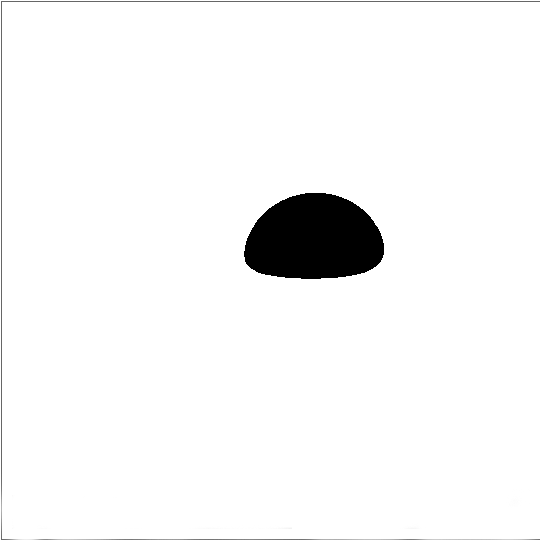}} &
    \fbox{\includegraphics[width=0.23\textwidth]{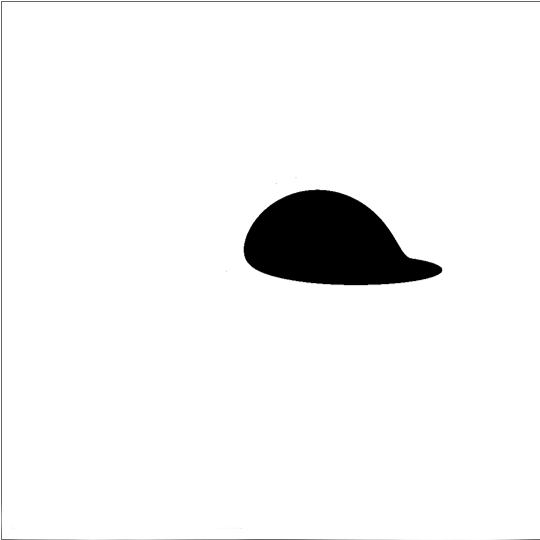}} \\
    
\end{tabular}
\caption{Inner shadows for different NUT parameters $n$.}
\label{10t}
\end{figure}

In Fig.\ref{fig:my_label}, the light blue dashed curve represents the direct image of the accretion disk, the green dashed curve is the lensed image of the accretion disk, as well as the black solid curve denotes the critical curve of the KTN black hole.
And, the extended inner shadow corresponds to the region enclosed by the dark blue and light blue solid curves. The part represented by the dark blue curve denotes the black hole horizon, while the part represented by the light blue curve indicates the region where light rays do not cross the equatorial plane.  
It can be seen that for an accretion disk located on the equatorial plane, with both geometrically and optically thin structure, the corresponding extended inner shadow of the KTN black hole exhibits significant differences from that of the Kerr black hole. Finally, we extracted the extended inner shadows of KTN black holes with different NUT charges and spin parameters and compared them with those of Kerr black holes, as shown in Fig.\ref{10t}.

As shown in Fig.\ref{10t}, when the observer is located at $\theta_o = 80^\circ$, the top row corresponds to $a = 0.2$, while the bottom row corresponds to $a = 0.94$. By comparing the inner shadows under different NUT parameters ($n = 0$, $0.1$, $0.3$), a clear evolutionary trend in their morphology can be observed.
In the absence of the NUT parameter ($n = 0$), the inner shadow remains symmetric, exhibiting a semicircle or semi-ellipse shape, consistent with the characteristics of traditional black holes. When $n = 0.1$, the inner shadow begins to show a rightward-tilting tendency, which becomes particularly pronounced under high-spin conditions, forming the rudimentary structure of a ``duck-bill cap'' shape.
As $n$ increases to $0.3$, the inner shadow adopts a typical right-tilted ``duck-bill cap'' shape. This asymmetric structure is most prominent under high-spin conditions ($a = 0.94$). This characteristic directly reflects the gravitomagnetic monopole effect in KTN spacetime.
With the ongoing improvements in EHT observational precision, it is expected that such a distinct inner shadow structure may be detected through future astronomical observations. If confirmed, this would provide key evidence for the existence of the gravitomagnetic monopole, holding significant implications for the refinement of gravitational theories.

\section{Conclusion } \label{sec4}

Within the framework of the optically thin accretion disk model, we focus on investigating the inner shadow and optical imaging features of rotating KTN black holes using the backward ray-tracing method. Specifically, we first perform a numerical evolution of the Hamiltonian canonical equations in the reference frame of a standard observer; Then, we construct circular and precessing orbital motions of the accretion flow, with the ISCO serving as the dividing line; and finally, the image of KTN black holes is obtained through stereographic projection techniques.

To explore the spacetime structure of the KTN black hole in depth, we present the observable images of optically thin accretion disks at both $\theta_o = 80^\circ$ and $\theta_o = 163^\circ$ in Fig.\ref{4o}.
It shows that the direct and lensed images remain distinguishable at $\theta_o=80^\circ$, while becoming indistinguishable at $\theta_o=163^\circ$. 
And, the left side of the photon ring appears significantly brighter than the right side, which is a result of the Doppler effect of the accretion disk - the image of the accretion flow on the side approaching the observer appears brighter, while that on the side moving away from the observer appears dimmer.
When fixed the NUT parameter, it is true that the photon ring on the left side of Fig.\ref{4b} appears flatter compared to Fig.\ref{4a}, and this flattening leads to a progressive deformation toward a ``D"-shaped profile. 
Increasing the spin parameter $a$ also deforms and shrinks the inner shadow, while a larger NUT parameter $n$ has the contrasting effect of enlarging both the inner shadow and the critical curve.
More interestingly, by comparing to Fig.\ref{4b}, the case with $n=0.3$ shows increased left-side brightness and an inner shadow whose stretched lower-right edge protrudes sharply beyond the critical curve, forming a ``duck-cap–like" profile, which is only found at $\theta_o = 80^\circ$.
Unlike any previously inner shadow, this morphology defines a novel type, which we call the black hole's ``extended inner shadow", which also be presented in Fig.\ref{10o}.
The later analysis in Fig.\ref{12i} shows that the dark region in the thin-disk image of a KTN black hole does not completely coincide with the event horizon, and we therefore term it the extended inner shadow. Notably, the tongue-shaped structure resembling a right-tilted “duck-cap” is not associated with the event horizon, as the corresponding photons do not cross the equatorial plane where the accretion disk-the light source-is located. As a result, this region receives no emission from the disk and appears dark. In contrast, for a Kerr black hole, photons in the corresponding region can cross the equatorial plane, leading to observable brightness in the image. This phenomenon demonstrates that the NUT charge, as a manifestation of the gravitomagnetic monopole, holds significance not only in altering the paths of light propagation but also in potentially providing a unique and irreplaceable probe.

Finally, we conclude that this distinctive inner shadow structure is expected to be tested by future high-precision astronomical observations, thereby providing new potential evidence for the existence of the black hole NUT charge. Nevertheless, the present study is primarily based on the thin-disk accretion model, and in subsequent work, we will further investigate the imaging properties of KTN black holes in the contexts of thick accretion disks and relativistic jets, with the aim of uncovering additional observable features associated with KTN black holes and offering a more comprehensive basis for probing their physical nature.

\section*{Acknowledgments}
This work is supported by the National Natural Science Foundation of China (GrantNo.12505059), and by the Sichuan Science and Technology Program (2024NSFSC1999).

\bibliographystyle{utphys}
\bibliography{note}

\end{document}